\documentclass[12pt, one column, a4paper] {article}

\usepackage[dvips]{epsfig}

\begin{document}

\title{Charged branes interactions via Kalb-Ramond field}

\author{F. A. Barone$^{a}$\footnote{e-mail: fbarone@unifei.edu.br} , F. E. Barone$^{b}$\footnote{e-mail: febarone@cbpf.br} and J. A. Helay\"el-Neto$^{b}$\footnote{e-mail: helayel@cbpf.br}\\
\small $^{a}$ ICE - Universidade Federal de Itajub\'a, Av. BPS 1303, Caixa Postal 50\\ 
\small 37500-903, Itajub\'a, MG, Brazil.\\
\small $^{b}$ Centro Brasileiro de Pesquisas F\'\i sicas, Rua Dr.\ Xavier Sigaud 150, Urca\\
\small  22290-180, Rio de Janeiro, RJ, Brazil}

\date{}

\maketitle





\begin{abstract}
Due to its versatility, the 2-form field has been employed to describe a multitude of scenarios that range from high energy to condensed matter physics. Pushing forward in this endeavor we study the interaction energy, intermediated by this kind of field, between branes in a variety of configurations. Also the so-called Cremmer-Scherk-Kalb-Ramond model, which consists of the electromagnetic field coupled to the Kalb-Ramond gauge potential, is considered. It turns out that these models exhibit a much richer class of sources than usually thought, able to intermediate novel forms of interactions in different scenarios.
\end{abstract}


\section{Introduction}

	The interaction between two static scalar charges intermediated by their coupling to a massless scalar field leads to an attractive coulombian interaction. A similar situation occurs for rank-2 symmetric tensors and static currents, whose interaction is intermediated by their coupling to the gravitational field. For two time-independent external vector sources, whose interaction is intermediated by the electromagnetic field, we have a repulsive coulombian force between equal charges. The attractive or repulsive nature of these forces are also present when we consider interactions between charges and multipoles distributions intermediated by these bosonic fields \cite{BaroneHidalgo,BaroneHidalgo2}. Nevertheless massive fields always intermediate short range interactions.
	
	On the other hand, it has been shown that the Dirac field can intermediate anisotropic interactions between point-like and time-independent external fermionic sources \cite{BaroneHidalgo} that are also long-range ones in the case of massless field. 
	
	Due to these results some questions can be raised in what concerns the influence of the spin, tensorial nature and mass of fields which can intermediate interactions between time-independent sources that simulate charges or multipole distributions. In order to have a better insight on these questions, in this work we pursue an investigation about some of these points in connection to the so called Kalb-Ramond field, the rank-2 skew-symmetric field, which is the gauge of strings \cite{KR}, that is coupled, in heterotic string theory \cite{SC}, to the Yang-Mills Chern-Simons three-form.
	
	It is worth mentioning that, recently, a lot of effort has been spent in trying to understand observable signatures of the Kalb-Ramond field, possibly due to an induced optical activity \cite{Gupta}, that would naturally occur in a braneworld scenario at low energies indicating a physics beyond the standard model. Besides, it has been employed to explain a vortex-boson duality in 3 dimensions that takes place in cuprate superconductivity\cite{PEREG}.
	
	 Specifically, in section (\ref{KR}), we investigate the interaction between two skew-symmetric rank-2 time-independent tensorial sources intermediated by their coupling to the Kalb-Ramond field. We consider a $3+1$ dimensional space-time and the currents are taken to be concentrated along two distinct parallel $D$-dimensional branes. We pay special attention to the case where the branes have zero dimension, that is, they are point-like. The current considered is interesting once it shows a point-like dipole aspect for the Kalb-Ramond field. The trivial fact that a 2-form field cannot be coupled to a point-like charge appears in a different notorious perspective and the results are extended to the interaction between two parallel linear charges distributions. We also argue why we can not have a true point-like dipole for the Kalb-Ramond field.
	 
	 In section (\ref{CSKR}), we study the so-called Cremmer-Scherk-Kalb-Ramond model \cite{CSKR,Lahiri}, which consists of the electromagnetic field coupled to the Kalb-Ramond one and can be taken as the generalization of the Chern-Simons model for $3+1$ dimensions. It appears in various scenarios, like supersymmetric models   \cite{Gates,Vilenkin,Baulieu}, cosmic strings \cite{Helayel}, cosmology \cite{Gasperini}, non commutative space-time  \cite{Amorin} and axions \cite{Lahiri2}. We choose the same external time-independent source coupled to the Kalb-Ramond field considered in the first model. For the electromagnetic sector we consider an external source composed by two uniform charges distributions along parallel branes. We show that, in spite of the considered model to be gauge invariant, when we have only electromagnetic point-like charges we get the Yukawa interaction between them. When we have only skew-symmetric point-like currents, the interaction between them is, in some circumstances, the one intermediated by a massive bosonic field, specifically, the interaction between two point-like charges and two point-like dipoles. By imposing appropriated restrictions to the skew-symmetric source, we can have only a Yukawa-type interaction or a dipole-dipole interaction for massive field. Similar analysis can be done when the branes are lines or planes.

\section{Rank-2 skew-symmetric currents}
\label{KR}

	In this section, we consider a $3+1$ space-time with the presence of two $D$-dimensional parallel branes. We shall denote by $d$ the perpendicular (spatial) dimensions to the branes in such a way to have $3=D+d$. Notice that $0\leq D,d< 3$ and when $D=0$ we have $d=3$ and the branes are points.	
	
	 We shall also denote the perpendicular and parallel spatial coordinates to the branes, respectively, by ${\bf x}_{\|}$ and ${\bf x}_{\perp}$ and use the metric signature $(+,-,-,...)$.

	Let us start by studying the model in $D+d+1$ dimensions given by the Lagrangian
\begin{equation}
\label{LKR}
{\cal L}_{KR}=\frac{1}{12}G^{\alpha\beta\gamma}G_{\alpha\beta\gamma}+\frac{1}{2}J^{\mu\nu}H_{\mu\nu}\ ,
\end{equation}
where $H_{\mu\nu}$ is the Kalb-Ramond field \cite{KR}, $J^{\mu\nu}$ is a skew-symmetric external source and $G^{\alpha\beta\gamma}$ is the field strength of the Kalb-Ramond field, given by
\begin{equation}
\label{defG}
G^{\alpha\beta\gamma}=\partial^{\alpha}H^{\beta\gamma}+\partial^{\beta}H^{\gamma\alpha}+\partial^{\gamma}H^{\alpha\beta}\ .
\end{equation}
The external source $J^{\mu\nu}$ has a vanishing divergence, $\partial_ {\mu}J^{\mu\nu}=0$.

	It is worthy mentioning that the Lagrangian (\ref{LKR}) exhibits the gauge invariance
\begin{equation}
\label{calibreKR}
H^{\alpha\beta} \to H^{\alpha\beta}+\partial^{\alpha}\xi^{\beta}-\partial^{\beta}\xi^{\alpha}\ ,
\end{equation}
where $\xi^{\alpha}$ is a four vector.

	With the standard Faddeev-Popov method for gauge fields, the generating functional for the Kalb-Ramond field can by written as the path integral
\begin{equation}
\label{funcionalKR}
{\cal Z}_{KR}=\int {\cal D}H\exp\Biggl[i\int d^{D+d+1}x\ {\cal L}-\frac{1}{2\alpha_{G}}(\partial_{\mu}H^{\mu\nu})(\partial_{\alpha}H^{\alpha}_{\ \nu})+\frac{1}{2}J_{\mu\nu}H^{\mu\nu}\Biggr]\ ,
\end{equation}
where $\alpha_{G}$ is a gauge parameter. Taking $\alpha_{G}=-1$ and performing the functional integral (\ref{funcionalKR}), as exposed in \cite{BaroneHelayel,Das}, we have
\begin{equation}
\label{funcionalKR2}
{\cal Z}_{KR}=\exp\Biggl[-\frac{i}{2}\int\ d^{D+d+1}x\ d^{D+d+1}y\ \frac{J_{\alpha\beta}(x)}{2}{\cal G}^{\alpha\beta,\mu\nu}(x,y)\frac{J_{\mu\nu}(y)}{2}\Biggr]\ ,
\end{equation}
where the Green's function is
\begin{equation}
\label{funcionalKR3}
{\cal G}_{\alpha\beta,\mu\nu}(x,y)=\int\frac{d^{D+d+1}p}{(2\pi)^{D+d+1}}\frac{1}{p^{2}}(\eta_{\alpha\mu}\eta_{\beta\nu}-\eta_{\alpha\nu}\eta_{\beta\mu})\exp[-ip(x-y)]\ .
\end{equation}

	From now on, we shall consider only time-independent sources, that is, $J_{\mu\nu}({\bf x})$. In this case the Lagrangian of the system does not exhibit explicit dependence on the time coordinate and, so, the corresponding functional generator can be written in the form \cite{Zee}
\begin{equation}
\label{funcionalgeral}
{\cal Z}_{KR}=\lim_{T\to\infty}\exp(-iET)\ ,
\end{equation}
where $E$ is the energy of the system and $T$ is the time interval.

	Comparing (\ref{funcionalKR2}) and (\ref{funcionalgeral}), we have
\begin{equation}
\label{energiaKRgeral}
E=\frac{1}{2T}\int\ d^{D+d+1}x\ d^{D+d+1}y\ \frac{J_{\alpha\beta}({\bf x})}{2}{\cal G}^{\alpha\beta,\mu\nu}(x,y)\frac{J_{\mu\nu}({\bf y})}{2}\ .
\end{equation}

	Now, let us search for an external source concentrated along two parallel $D$-dimensional branes which resembles the presence of charge or multipole distributions for the Kalb-Ramond field.

	For the electromagnetic field, it can be shown that \cite{BaroneHidalgo} static and uniform distributions of electric charges along branes can be described by external four-vector sources proportional to Dirac delta functions concentrated along the branes. Due to its symmetries, it is well known that the Kalb-Ramond field is not consistent with sources proportional to Dirac delta functions concentrated at points of space, as shall be exposed at the end of this section. The most simple skew-symmetric external time-independent source with vanishing four-divergence and concentrated along a brane with arbitrary dimension, (including dimension zero, that is, points) is given by
\begin{equation}
\label{correnteKRgeral}
J^{\mu\nu}({\bf x}_{\perp})=\epsilon^{\mu\nu\alpha\beta}V_{\alpha}\partial_{\beta}\Bigl(\delta^{d}({\bf x}_{\perp}-{\bf a})\Bigr)\ ,
\end{equation}
where $\epsilon^{\mu\nu\alpha\beta}$ is the Levi-Civita tensor with $\epsilon^{0123}=1$, $V^{\alpha}$ is a pseudo four-vector taken to be constant and uniform in the reference frame where the calculations are performed and ${\bf a}$ is a $d$-dimensional spatial vector, ${\bf a}=(a^{1},...,a^{d})$.

The source (\ref{correnteKRgeral}) is concentrated along the $D$-dimensional brane located at ${\bf a}$. It is important to notice that we are restricted to a 3+1 space-time and $d$ can assume the values 3, 2 or 1, where the brane becomes a point, a line and a plane, respectively.

	Although we are considering only static configurations, note that, for a given $d$, the source (\ref{correnteKRgeral}) can be taken as a special case of an expression valid also to moving branes. For instance, in the case of a point-like brane with world line $a^{\mu}(\tau)$ and four velocity $u^{\mu}(\tau)=da^{\mu}(\tau)/d\tau$, where $\tau$ stands for the proper time, we have the skew-symmetric source
\begin{equation}
\label{correntemov}
J^{\mu\nu}(x)=\int d\tau \epsilon^{\mu\nu\alpha\beta}V_{\alpha}(\tau)\Bigl(\eta_{\beta\lambda}-u_{\beta}(\tau)u_{\lambda}(\tau)\Bigr)\partial^{\lambda}\Bigl[\delta^{4}\Bigl(x^{\gamma}-a^{\gamma}(\tau)\Bigr)\Bigr]\ ,   
\end{equation}
where $\delta^{4}\Bigl(x^{\gamma}-a^{\gamma}(\tau)\Bigr)=\delta^{3}({\bf x}-{\bf a}(\tau))\delta(x^{0}-a^{0}(\tau))$.
If $V^{\mu}$ is constant and the brane is fixed, that is, $a^{0}=x^{0}=\tau$, ${\bf a}=constant$ and $u^{\mu}=\eta^{\mu0}$, the source (\ref{correntemov}) reduces to (\ref{correnteKRgeral}) with $d=3$, as follows,
\begin{eqnarray}\nonumber
J^{\mu\nu}(x)&=&\int d\tau \epsilon^{\mu\nu\alpha\beta}V_{\alpha}\Bigl(\eta_{\beta\lambda}-
\eta_{\beta0}\eta_{\lambda0}\Bigr)\partial^{\lambda}\Bigl[\delta^{3}({\bf x}-{\bf a})\delta(x^{0}-\tau)\Bigr]\cr\cr
&=&\int d\tau \epsilon^{\mu\nu\alpha\beta}V_{\alpha}\Bigl(\eta_{\beta\lambda}-
\eta_{\beta0}\eta_{\lambda0}\Bigr)\partial^{\lambda}\Bigl[\delta^{3}({\bf x}-{\bf a})\Bigr]\delta(x^{0}-\tau)\cr\cr
&=&\int d\tau \epsilon^{\mu\nu\alpha\beta}V_{\alpha}\eta_{\beta\lambda}\partial^{\lambda}\Bigl[\delta^{3}({\bf x}-{\bf a})\Bigr]\delta(x^{0}-\tau)\cr\cr
&=&\epsilon^{\mu\nu\alpha\beta}V_{\alpha}\eta_{\beta\lambda}\partial^{\lambda}\Bigl[\delta^{3}({\bf x}-{\bf a})\Bigr],
\end{eqnarray}
where in the second and third lines we used the fact that when $\lambda=0$ the integral vanishes

	Once the source (\ref{correnteKRgeral}) is a skew-symmetric tensor, its components $i0$, $i=1,2,3$, build up the polar vector
\begin{eqnarray}
J^{i0}({\bf x}_{\perp})&=&\epsilon^{0i\alpha\beta}V_{\beta}\partial_{\alpha}\delta^{d}({\bf x}_{\perp}-{\bf a})=\Bigl[{\bf V}\times\Bigl({\bf\nabla}\delta^{d}({\bf x}_{\perp}-{\bf a})\Bigr)\Bigr]^{i}\ ,
\end{eqnarray}
which has vanishing divergence, and	with the spatial part we may construct an axial vector as follows,
\begin{equation}
\epsilon^{ijk}J^{ij}({\bf x}_{\perp})=2V^{0}\Bigl({\bf\nabla}\delta^{d}({\bf x}_{\perp}-{\bf a})\Bigr)^{k}\ .
\end{equation}

	Now let us take an external source composed by two terms, each one in the form (\ref{correnteKRgeral}), but concentrated along different (parallel) branes,  
\begin{equation}
\label{zxc1}
J^{\mu\nu}({\bf x}_{\perp})=\epsilon^{\mu\nu\alpha\beta}\partial_{\beta}\Bigl[V_{\alpha}\delta^{d}({\bf x}_{\perp}-{\bf a}_{1})+W_{\alpha}\delta^{d}({\bf x}_{\perp}-{\bf a}_{2})\Bigr]\ .
\end{equation}
In the equation above, $V_{\alpha}$ and $W_{\alpha}$ are two pseudo four-vectors and ${\bf a}_{1}$ and ${\bf a}_{2}$ are two distinct spatial $d$-dimensional vectors.

	Substituting (\ref{zxc1}) in (\ref{energiaKRgeral}), performing two integrations by parts and discarding the terms due to the self-interaction of a brane with itself (which does not contribute to the force between the branes) we have
\begin{eqnarray}
\label{zxc20}
E=-\frac{1}{2T}\int d^{d+D+1}x \int d^{d+D+1}y \int\frac{d^{d+D+1}p}{(2\pi)^{d+D+1}}\epsilon^{\mu\nu\alpha\beta}\epsilon^{\lambda\rho\sigma\tau}\Bigl(\partial_{\alpha(x)}\partial_{\sigma(x)}{\cal G}_{\mu\nu,\lambda\rho}(x,y)\Bigr)\cr\cr
\Biggl[\frac{V_{\beta}W_{\tau}}{4}\delta^{d}({\bf x}_{\perp}-{\bf a}_{1})\delta^{d}({\bf y}_{\perp}-{\bf a}_{2})+\frac{V_{\tau}W_{\beta}}{4}\delta^{d}({\bf x}_{\perp}-{\bf a}_{2})\delta^{d}({\bf y}_{\perp}-{\bf a}_{1})\Biggr]\ .
\end{eqnarray}

	Using the fact that ${\cal G}_{\mu\nu,\lambda\rho}(x,y)={\cal G}_{\mu\nu,\lambda\rho}(y,x)$ and the Fourier representation (\ref{funcionalKR3}), performing a change of variables and, in the following order, the integrals $dx^{0}dp^{0}dy^{0}d{\bf x}_{\|}d{\bf p}_{\|}d{\bf y}_{\|}d{\bf x}_{\perp}d{\bf y}_{\perp}$, defining the difference vector ${\bf a}={\bf a}_{1}-{\bf a}_{2}$, the $D$-dimensional area of a brane $L^{D}=\int d^{D}{\bf y}$ and using the fact that $T=\int dy^{0}$, we can write the energy (\ref{zxc20}) in the form
\begin{eqnarray}
\label{zxc2}
E=-\frac{1}{4}L^{D}\int\frac{d^{d}{\bf p}_{\perp}}{(2\pi)^{d}}\epsilon^{\mu\nu\alpha\beta}\epsilon^{\lambda\rho\sigma\tau}\frac{1}{2}(V_{\beta}W_{\tau}+V_{\tau}W_{\beta})(\eta_{\mu\lambda}\eta_{\nu\rho}-\eta_{\mu\rho}\eta_{\nu\lambda})\frac{p_{\alpha}p_{\sigma}}{{\bf p}_{\perp}^{2}}\exp(i{\bf p}_{\perp}\cdot{\bf a})\ ,\cr
0<\alpha,\sigma\leq d\ ,
\end{eqnarray}

	With the aid of the identity $\epsilon^{\mu\nu\alpha\beta}\epsilon_{\mu\nu}^{\ \ \ \sigma\tau}=-2(\eta^{\alpha\sigma}\eta^{\beta\tau}-\eta^{\alpha\tau}\eta^{\beta\sigma})$, Eq. (\ref{zxc2}) leads to
\begin{equation}
\label{zxc3}
{\cal E}=\frac{E}{L^{D}}=-\int\frac{d^{d}{\bf p}_{\perp}}{(2\pi)^{d}}[(V^{\tau}W_{\tau}){\bf p}_{\perp}^{2}+({\bf V}\cdot{\bf p}_{\perp})({\bf W}\cdot{\bf p}_{\perp})]\frac{1}{{\bf p}_{\perp}^{2}}\exp(i{\bf p}_{\perp}\cdot{\bf a})\ ,
\end{equation}
where we have defined the energy per unity of brane's area, ${\cal E}=E/L^{D}$.

	In order to calculate the integral (\ref{zxc3}), we define the operator ${\bf\nabla}_{a}=(\partial/\partial a^{1}_{12},...,\partial/\partial a^{d}_{12})$ and rewrite Eq. (\ref{zxc3}) in the form
\begin{equation}
{\cal E}=[(V^{\tau}W_{\tau}){\bf\nabla}_{a}^{2}+({\bf V}\cdot{\bf\nabla}_{a})({\bf W}\cdot{\bf\nabla}_{a})]\int\frac{d^{d}{\bf p}_{\perp}}{(2\pi)^{d}}\frac{1}{{\bf p}_{\perp}^{2}}\exp(i{\bf p}_{\perp}\cdot{\bf a})\ .
\end{equation}

	The integral above is calculated in \cite{BaroneHidalgo}. For this task, it is necessary to consider the case where $d=2$ and $d\not=2$. For $d\not=2$ we have 
\begin{equation}
\label{zxc4}
\int\frac{d^{d}{\bf p}_{\perp}}{(2\pi)^{d}}\frac{1}{{\bf p}_{\perp}^{2}}\exp(i{\bf p}_{\perp}\cdot{\bf a})=\frac{1}{(2\pi)^{d/2}}2^{(d/2)-2}\Gamma\Bigl((d/2)-1\Bigr)a^{2-d}\ ,\ d\not=2\ ,
\end{equation}
where $\Gamma$ stands for the Gamma function.

	For $d=2$, we proceed as in references \cite{BaroneHidalgo,BaroneHidalgo2}, and introduce a mass parameter, as a regulator parameter, and an arbitrary positive constant $a_{0}$ with dimension of length, which does not contribute to the force between the branes, as follows
\begin{eqnarray}
\label{zxc21}
\lim_{m\to0}\Biggl(\int\frac{d^{2}{\bf p}_{\perp}}{(2\pi)^{2}}\frac{1}{{\bf p}_{\perp}^{2}+m^{2}}\exp(i{\bf p}_{\perp}\cdot{\bf a})\Biggr)&=&\frac{1}{2\pi}\lim_{m\to0}K_{0}(ma)\cr\cr
&=&\frac{1}{2\pi}\lim_{m\to0}\Biggl(-\gamma-\ln\frac{ma}{2}+\ln a_{0}-\ln a_{0}\Biggr)\ .
\end{eqnarray}
In the equation above, $\gamma$ is the Euler constant, $K_{0}(x)$ is the $K$-Bessel function of order 0 and we have used the fact that \cite{Arfken} $K_{0}(x)\stackrel{x\to0}{\longrightarrow}-\gamma-\ln(x/2)$.

	Separating the $a$-independent terms at the right-hand-side (RHS) of (\ref{zxc21}), 
\begin{equation}
\label{zxc22}
\lim_{m\to0}\Biggl(\int\frac{d^{2}{\bf p}_{\perp}}{(2\pi)^{2}}\frac{1}{{\bf p}_{\perp}^{2}+m^{2}}\exp(i{\bf p}_{\perp}\cdot{\bf a})\Biggr)=-\frac{1}{2\pi}\ln\frac{a}{a_{0}}-\frac{1}{2\pi}\lim_{m\to0}\Biggl(\gamma+\ln\frac{ma_{0}}{2}\Biggr)\ .
\end{equation}
In spite of being divergent, the second term at the RHS of (\ref{zxc22}) is $a$-independent and does not contribute to the force between the branes. So, it shall be discarded from now on.

	Substituting (\ref{zxc4}) or (\ref{zxc22}) in (\ref{zxc3}), we are led, after some manipulations, to the same result
\begin{equation}
\label{KRdipolo}
{\cal E}=-\frac{1}{(2\pi)^{d/2}}2^{(d/2)-1}\Gamma(d/2)\frac{1}{a^{d}}
\Bigl[({\bf V}_{\perp}\cdot{\bf W}_{\perp})-d({\bf V}\cdot{\hat a})({\bf W}\cdot{\hat a})\Bigr]\ , 
\end{equation}
where ${\hat a}={\bf a}_{12}/|{\bf a}_{12}|$ and $a=|{\bf a}_{12}|$.

	It is interesting to notice that result (\ref{KRdipolo}) is the same as the one obtained by an uniform dipole distributions along branes placed at ${\bf a}_{1}$ and ${\bf a}_{2}$, but with an overall minus sign \cite{BaroneHidalgo,BaroneHidalgo2}.


	In spite of the four vectors $V^{\mu}$ and $W^{\mu}$ in (\ref{zxc1}) can have perpendicular as well as parallel components to the branes, only their perpendicular components contribute to the energy (\ref{KRdipolo}).

	As shall be discussed, the Kalb-Ramond field can not couple with a true point-like dipole, but the source (\ref{zxc1}) can be taken as being concentrated at two points. When $D=0$, the branes are points and, once the space-time has $3+1$ dimensions, we have $d=3$, ${\cal E}=E$, ${\bf V}_{\perp}={\bf V}$ and ${\bf W}_{\perp}={\bf W}$. In this situation we can write
\begin{equation}
\label{zxc5}
E(d=3)=-\frac{1}{4\pi}\frac{1}{a^{3}}\Bigl[({\bf V}\cdot{\bf W})-3({\bf V}\cdot{\hat a})({\bf W}\cdot{\hat a})\Bigr]\ .
\end{equation}

	The energy (\ref{zxc5}) is the same as the one obtained for the interaction between two electric dipoles ${\bf V}$ and ${\bf W}$ placed at the points ${\bf a}_{1}$ and ${\bf a}_{2}$, but with an overall minus sign, as expected for a rank-2 field \cite{Gravitacao}. A similar situation occurs for the scalar field where we have an overall minus sign in comparing with the electromagnetic case.
	
	 So the source (\ref{correnteKRgeral}) can be interpreted as a kind of uniform distribution of dipoles for the Kalb-Ramond field along the branes which can be generalized to the case when the branes are points.
	 
	For completeness, to conclude this section, we would like emphasize the well known fact that the Kalb-Ramond field is not compatible with point-like time-independent external sources with vanishing four-divergences which describe true point-like charges or point-like dipoles, but it is compatible with external sources which describe uniform distributions of charges or dipoles along $D$-dimensional branes if $D\geq1$. Actually, being a 2-form gauge potential, the Kalb-Ramond field can, consistently, describe the interaction between extended 1-dimensional objects (strings) whose dynamics is associated to world surfaces, rather than world lines.	This fact can be seen by considering the Lagrangian (\ref{LKR}) in a space-time with arbitrary space dimensions and the external source
\begin{equation}
\label{KRJmonopolo}
J^{\mu\nu}({\bf x}_{\perp})=V^{\mu\nu}\delta^{d}({\bf x}_{\perp}-{\bf a}_{1})+W^{\alpha\beta}\delta^{d}({\bf x}_{\perp}-{\bf a}_{2})\ ,
\end{equation}
where $V^{\mu\nu}$ and $W^{\mu\nu}$ are rank-2 constant and uniform skew-symmetric tensors satisfying $V^{\mu\nu}=W^{\mu\nu}=0$ for $\mu=1,...,d$ (or $\nu=1,...,d$), what assures a vanishing four divergence for (\ref{KRJmonopolo}).

	Notice that, if $D=0$, the branes are points and the space becomes $d$-dimensional, so $V^{\mu\nu}=W^{\mu\nu}=0$ for $\mu,\nu\not=0$. But, once $V^{\mu\nu}$ and $W^{\mu\nu}$ are skew-symmetric tensors, all their components must be zero.	

	Following similar steps which led to (\ref{KRdipolo}), the source (\ref{KRJmonopolo}) gives the interaction energy between the branes located at ${\bf a}_{1}$ and ${\bf a}_{2}$ per unity of the brane's area 
\begin{eqnarray}
\label{EKRmonopolo}
{\cal E}&=&-\frac{V^{\mu\nu}W_{\mu\nu}}{2}\frac{1}{(2\pi)^{d/2}}2^{(d/2)-2}\Gamma\Bigl((d/2)-1\Bigr)a^{2-d}\ \ ,\ \ d\not=2\cr\cr
&=&\frac{V^{\mu\nu}W_{\mu\nu}}{2}\frac{1}{2\pi}\ln\frac{a}{a_{0}}\ \ ,\ \ d=2\ .
\end{eqnarray}

	Each term in (\ref{KRJmonopolo}) can be interpreted as a charge density for the Kalb-Ramond field. In the case where $V^{\mu\nu}W_{\mu\nu}>0$ the interaction energy (\ref{EKRmonopolo}) gives an attractive force between the branes, which means that the Kalb-Ramond field has an opposite character in comparing with the electromagnetic field in what concerns the sign of the force between charges distributions. We can say that, for the Kalb-Ramond field, distributions of charges with equal sign attract one another as expected for a 2-rank tensor.
	
	The same analysis could be done for the rank-2 source
\begin{equation}
\label{KRJmonopoloverdadeiro}
J^{\mu\nu}({\bf x}_{\perp})=V^{\mu\nu}\mbox{v}^{\rho}\partial_{\rho}\delta^{d}({\bf x}_{\perp}-{\bf a}_{1})+W^{\alpha\beta}\mbox{w}^{\gamma}\partial_{\gamma}\delta^{d}({\bf x}_{\perp}-{\bf a}_{2})\ ,
\end{equation}
which describes true uniform dipole distributions along parallel branes for the Kalb-Ramond field. In the above source, $\mbox{v}^{\rho}$ and $\mbox{w}^{\rho}$ are two four-vectors and we must have $d\geq1$, so that the branes can not be point-like.

\section{Cremmer-Scherk-Kalb-Ramond source}
\label{CSKR}

	In this section, we study the so-called Cremmer-Scherk-Kalb-Ramond model, whose Lagrangian is given by \cite{CSKR}
\begin{equation}
\label{LCSKR}
{\cal L}_{CSKR}=\frac{1}{12}G^{\alpha\beta\gamma}G_{\alpha\beta\gamma}-\frac{1}{4}F^{\mu\nu}F_{\mu\nu}-\frac{\mu}{4}\epsilon^{\mu\nu\alpha\beta}H_{\mu\nu}F_{\alpha\beta}+\frac{1}{2}J^{\mu\nu}H_{\mu\nu}+J^{\mu}A_{\nu}\ ,
\end{equation}
where $A^{\mu}$ is the electromagnetic field, $H^{\mu\nu}$ is the Kalb-Ramond field, $G^{\alpha\beta\gamma}$ is the field strength for $H^{\mu\nu}$ defined in (\ref{defG}), $F^{\mu\nu}=\partial^{\mu}A^{\nu}-\partial^{\nu}A^{\mu}$ is the field strength for $A^{\mu}$ and $\mu$ is the coupling constant among the Kalb-Ramond and vector fields.
  
	It can be shown that the Lagrangian (\ref{LCSKR}) exhibits the gauge symmetry (\ref{calibreKR}) for the Kalb-Ramond field and the usual gauge symmetry, $A^{\mu}\to A^{\mu}+\partial^{\mu}\Lambda$, for the electromagnetic field. The external sources $J^{\mu\nu}$ and $J^{\mu}$ must have vanishing divergence as dictated by the respective gauge symmetries they are associated to. The Lagrangian (\ref{LCSKR}) can be seen as a generalization to $3+1$ dimensions of the Chern-Simons model \cite{Lahiri}.
		
	Using standard Fadeev-Popov methods for gauge fields, the generating functional for the Lagrangian (\ref{LCSKR}) can be written in the form
\begin{eqnarray}
\label{asd1}
{\cal Z}_{CSKR}=\int{\cal D}H{\cal D}A\exp\Biggl[i\int d^{4}x\ \frac{1}{12}G^{\alpha\beta\gamma}G_{\alpha\beta\gamma}-\frac{1}{4}F^{\mu\nu}F_{\mu\nu}-\frac{\mu}{4}\epsilon^{\mu\nu\alpha\beta}H_{\mu\nu}F_{\alpha\beta}\cr\cr
-\frac{1}{2\alpha_{G}}(\partial_{\mu}H^{\mu\nu})(\partial_{\alpha}H^{\alpha}_{\ \nu})-\frac{1}{2\beta_{G}}(\partial_{\mu}A^{\mu})(\partial_{\nu}A^{\nu})+\frac{1}{2}J^{\mu\nu}H_{\mu\nu}+J^{\mu}A_{\mu}\Biggr]\ ,
\end{eqnarray}
where $\alpha_{G}$ and $\beta_{G}$ are gauge parameters.

	Choosing the gauges with $\alpha_{G}=-1$ and $\beta_{G}=1$, and performing some simple manipulations, the functional (\ref{asd1}) can be reduced to	
\begin{eqnarray}
\label{asd2}
{\cal Z}_{CSKR}=\int{\cal D}H{\cal D}A\exp\Biggl[i\int d^{4}x H^{\alpha\beta}\Biggl(-\frac{1}{8}(\eta_{\alpha\mu}\eta_{\beta\nu}-\eta_{\alpha\nu}\eta_{\beta\mu})\partial_{\tau}\partial^{\tau}\Biggr)H^{\mu\nu}+\frac{1}{2}J^{\mu\nu}H_{\mu\nu}\cr\cr
+\frac{1}{2}A^{\gamma}(\eta_{\gamma\lambda}\partial_{\tau}\partial^{\tau})A^{\lambda}+J^{\mu}A_{\mu}
-\frac{\mu}{4} H^{\alpha\beta}\epsilon_{\alpha\beta\tau\lambda}\partial^{\tau}A^{\lambda}
-\frac{\mu}{4} A^{\gamma}\epsilon_{\gamma\tau\mu\nu}\partial^{\tau}H^{\mu\nu}\Biggr]\ .
\end{eqnarray}

	Defining the field ${\cal X}^{\mu\nu,\lambda}$ and the current $J^{\mu\nu,\lambda}$, with matrix structures, as follows
\begin{equation}
{\cal X}^{\mu\nu,\lambda}=
  \left(
         \begin{array}{rrrr}
          H^{\mu\nu} \cr
          A^{\lambda}\ 
         \end{array}
  \right)
,\ \ \ 
J^{\mu\nu,\lambda}=
  \left(
         \begin{array}{rrrr}
          J^{\mu\nu}/2 \cr
          J^{\lambda}\ \ 
          \end{array}
  \right)
\end{equation}
and the differential operator
\begin{equation}
\label{defK}
K_{\alpha\beta,\gamma;\mu\nu,\lambda}(x)=
  \left(
         \begin{array}{cccc}
 -\frac{1}{4}(\eta_{\alpha\mu}\eta_{\beta\nu}-\eta_{\alpha\nu}\eta_{\beta\mu})\partial_{\tau}\partial^{\tau}
  & -\frac{\mu}{2}\epsilon_{\alpha\beta\tau\lambda}\partial^{\tau}    \cr\cr
	  -\frac{\mu}{2}\epsilon_{\gamma\tau\mu\nu}\partial^{\tau}
	& \eta_{\gamma\lambda}\partial_{\tau}\partial^{\tau}                                                           
         \end{array}
  \right)\ ,
\end{equation}
we can rewrite the generating functional (\ref{asd2}) in the form
\begin{eqnarray}
\label{asd3}
{\cal Z}_{CSKR}=\int{\cal D}{\cal X}\exp\Biggl[\frac{i}{2}\int d^{4}x\ {\cal X}^{\dagger\ \alpha\beta,\gamma}K_{\alpha\beta,\gamma;\mu\nu,\lambda}(x){\cal X}^{\mu\nu,\lambda}+J^{\mu\nu,\lambda}{\cal X}_{\mu\nu,\lambda}\Biggr]\ ,
\end{eqnarray}
which exhibits a Gaussian form in the field ${\cal X}^{\alpha\beta,\gamma}$.

	Computing the Gaussian integral (\ref{asd3}), we get
\begin{equation}
\label{asd31}
{\cal Z}_{CSKR}=\exp\Biggl[-\frac{i}{2}\int d^{4}xd^{4}y J^{\dagger\ \alpha\beta,\gamma}(x)K^{-1}_{\alpha\beta,\gamma;\mu\nu,\lambda}(x,y)J^{\mu\nu,\lambda}(y)\Biggr]\ ,
\end{equation}
where $K^{-1}_{\alpha\beta,\gamma;\mu\nu,\lambda}(x,y)$ is the inverse of the operator (\ref{defK}) in the sense that
\begin{equation}
\label{asd4}
K_{\alpha\beta,\gamma;\mu\nu,\lambda}(x)K^{-1\ \mu\nu,\lambda}_{\ \ \ \ \ \ \ \ \sigma\rho,\theta}(x,y)=
  \left(
         \begin{array}{cccc}
   \frac{1}{2}(\eta_{\alpha\sigma}\eta_{\beta\rho}-\eta_{\alpha\rho}\eta_{\beta\sigma})
 & 0  \cr
   0  
 & \eta_{\gamma\theta}
         \end{array}
  \right)
\delta^{4}(x-y)\ .
\end{equation}

	The Green's function $K^{-1}_{\alpha\beta,\gamma;\mu\nu,\lambda}(x,y)$ can be calculated as a Fourier integral; the result is
\begin{equation}
\label{defK-1}
K^{-1\ \mu\nu,\lambda}_{\ \ \ \ \ \ \ \ \sigma\rho,\theta}(x,y)=
\int\frac{d^{4}p}{(2\pi)^{4}}
  \left(
         \begin{array}{cccc}
   \tilde{\cal A}^{\mu\nu}_{\ \ \sigma\rho}(p)
 & \tilde{\cal B}^{\mu\nu}_{\ \ \theta}(p)  \cr\cr
   \tilde{\cal C}^{\lambda}_{\ \sigma\rho}(p)  
 & \tilde{\cal D}^{\lambda}_{\ \theta}(p)
         \end{array}
  \right)
\exp[-i(x-y)]\ ,
\end{equation}
where
\begin{eqnarray}
\label{defABCD}
\tilde{\cal A}^{\mu\nu}_{\ \ \sigma\rho}(p)&=&\frac{2}{p^{2}-\mu^{2}}\Biggl[\eta^{\mu}_{\ \ [\sigma}\eta^{\nu}_{\ \ \rho]}-\frac{2\mu^{2}}{p^{2}}\eta^{[\mu}_{\ \ [\sigma}\frac{p^{\nu]}p_{\rho]}}{p^{2}}\Biggr]\ ,  \cr\cr
\tilde{\cal B}^{\mu\nu}_{\ \ \theta}(p)&=&i\mu\frac{1}{p^{2}}\frac{1}{p^{2}-\mu^{2}}\epsilon^{\mu\nu}_{\ \  \tau\theta}p^{\tau}\ ,\cr\cr
\tilde{\cal C}^{\lambda}_{\ \sigma\rho}(p)&=&i\mu\frac{1}{p^{2}}\frac{1}{p^{2}-\mu^{2}}\epsilon^{\lambda\tau}_{\ \ \sigma\rho}p^{\tau}\ ,    \cr\cr
\tilde{\cal D}^{\lambda}_{\ \theta}(p)&=&-\frac{1}{p^{2}-\mu^{2}}\Biggl[\eta^{\lambda;}_{\ \ \theta} -\frac{\mu^{2}}{p^{2}}\frac{p^{\lambda}p_{\theta}}{p^{2}}\Biggr]\ .
\end{eqnarray}

	Once the sources $J^{\mu\nu}$ and $J^{\mu}$ do not depend on time, we have that ${\cal Z}_{CSKR}=\exp(-iET)$, where $E$ is the energy of the system. So, by using (\ref{asd31}) 
\begin{equation}
\label{asd5}
E=\frac{1}{2T}\int d^{4}xd^{4}y J^{\dagger\ \alpha\beta,\gamma}({\bf x})K^{-1}_{\alpha\beta,\gamma;\mu\nu,\lambda}(x,y)J^{\mu\nu,\lambda}({\bf y})\ .
\end{equation}

	With the aid of (\ref{defABCD}) and (\ref{defK-1}), and performing similar steps which led to Eq. (\ref{zxc2}), it can be shown that the energy (\ref{asd5}) is the sum of the following four parts
\begin{eqnarray}
\label{E=E1+E2+E3+E4}
E_{1}&=&
\frac{1}{2}\int d^{3}{\bf x}d^{3}{\bf y}\int\frac{d^{3}{\bf p}}{(2\pi)^{3}}\frac{J^{\alpha\beta}({\bf x})}{2}\tilde{\cal A}_{\alpha\beta;\mu\nu}(p)\frac{J^{\mu\nu}({\bf y})}{2}\exp[i{\bf p}\cdot({\bf x}-{\bf y})]\ ,\cr\cr
E_{2}&=&
\frac{1}{2}\int d^{3}{\bf x}d^{3}{\bf y}\int\frac{d^{3}{\bf p}}{(2\pi)^{3}}\frac{J^{\alpha\beta}({\bf x})}{2}\tilde{\cal B}_{\alpha\beta;\lambda}(p)J^{\lambda}({\bf y})\exp[i{\bf p}\cdot({\bf x}-{\bf y})]\ ,\cr\cr
E_{3}&=&
\frac{1}{2}\int d^{3}{\bf x}d^{3}{\bf y}\int\frac{d^{3}{\bf p}}{(2\pi)^{3}}J^{\gamma}({\bf x})\tilde{\cal C}_{\gamma;\mu\nu}(p)\frac{J^{\mu\nu}({\bf y})}{2}\exp[i{\bf p}\cdot({\bf x}-{\bf y})]\ ,\cr\cr
E_{4}&=&
\frac{1}{2}\int d^{3}{\bf x}d^{3}{\bf y}\int\frac{d^{3}{\bf p}}{(2\pi)^{3}}J^{\gamma}({\bf x})\tilde{\cal D}_{\gamma;\lambda}(p)J^{\lambda}({\bf y})\exp[i{\bf p}\cdot({\bf x}-{\bf y})] .
\end{eqnarray}

	Now, in order to have a better insight on the meaning of time-independent sources distributions for the Cremmer-Scherk-Kalb-Ramond model, let us start by considering sources that, in the reference frame we are performing the calculations, are distributed along $D$-dimensional parallel branes and specify our analysis to the following external sources
\begin{eqnarray}
\label{correntetotal}
J^{\mu\nu}({\bf x}_{\perp})&=&\epsilon^{\mu\nu\alpha\beta}\Bigl[V_{\beta}\partial_{\alpha}\delta^{d}({\bf x}_{\perp}-{\bf a}_{1})+W_{\beta}\partial_{\alpha}\delta^{d}({\bf x}_{\perp}-{\bf a}_{2})\Bigr]\cr
J^{\mu}({\bf x}_{\perp})&=&v^{\mu}\delta^{d}({\bf x}_{\perp}-{\bf a}_{1})+w^{\mu}\delta^{d}({\bf x}_{\perp}-{\bf a}_{2})\ .
\end{eqnarray}
where $v^{\mu}$ and $w^{\mu}$ are constant and uniform four vectors that, as discussed in reference \cite{BaroneHidalgo}, are restricted to the conditions $v^{1}=w^{1}=..=v^{d}=w^{d}=0$ in order to assure gauge invariance for the vector sector.

 	The source $J^{\mu\nu}$ above is the same as the one given in Eq. (\ref{zxc1}) and discussed in the previous section. Its role for the model (\ref{LCSKR}) shall be clearer at the end of the section. As pointed out in references \cite{BaroneHidalgo,BaroneHidalgo2}, if we take $v^{\mu},w^{\mu}\sim\eta^{\mu0}$, the external source $J^{\mu}$ defined in Eq. (\ref{correntetotal}) describes two uniform electric charges distributions along the branes placed at ${\bf a}_{1}$ and ${\bf a}_{2}$. So we have an external source for the Kalb-Ramond and Maxwell fields at the branes ${\bf a}_{1}$ and ${\bf a}_{2}$, all of them being time-independent.

	The space-time is $3+1$ dimensional, so we must have $d=1,2,3$, where the branes are points, lines and planes, respectively.
	
	The contribution $E_{1}$ in (\ref{E=E1+E2+E3+E4}) can be calculated following similar steps we have used in the previous section,
\begin{eqnarray}
\label{asd7}
E_{1}&=&L^{D}[(V_{\lambda}W^{\lambda}{\bf\nabla}_{a}^{2}+({\bf V}\cdot{\bf\nabla}_{a})({\bf W}\cdot{\bf\nabla}_{a})]\int\frac{d^{d}{\bf p}_{\perp}}{(2\pi)^{d}}\frac{1}{p_{\perp}^{2}+\mu^{2}}\exp(i{\bf p}_{\perp}\cdot{\bf a})\cr\cr
&=&L^{D}[(V_{\lambda}W^{\lambda}{\bf\nabla}_{a}^{2}+({\bf V}\cdot{\bf\nabla}_{a})({\bf W}\cdot{\bf\nabla}_{a})]
\frac{1}{(2\pi)^{d/2}}\mu^{d-2}(\mu a)^{1-(d/2)}K_{(d/2)-1}(\mu a)\ ,
\end{eqnarray}
where we defined ${\bf a}={\bf a}_{1}-{\bf a}_{2}$, assumed that $\mu\geq0$, used the fact that \cite{BaroneHidalgo}
\begin{equation}
\label{integral}
\int\frac{d^{d}{\bf p}}{(2\pi)^{d}}\frac{1}{p^{2}+\mu^{2}}\exp(i{\bf p}\cdot{\bf a})=\frac{1}{(2\pi)^{d/2}}\mu^{d-2}(\mu a)^{1-(d/2)}K_{(d/2)-1}(\mu a)
\end{equation}
and discarded $a$-independent terms due to self interactions.

	The quantity $E_{1}$ is a contribution to the energy due solely to the interaction between the two parts of the tensor source $J^{\mu\nu}$ and does not involve the vector source $J^{\mu}$.	
		
	The contribution $E_{4}$ can be obtained with the aid of Eq's (\ref{defABCD}), (\ref{E=E1+E2+E3+E4}), (\ref{correntetotal}) and (\ref{integral}), and from the results presented in \cite{BaroneHidalgo},
\begin{equation}
\label{asd8}
E_{4}=L^{D}\frac{v_{\lambda}w^{\lambda}}{(2\pi)^{d/2}}\mu^{d-2}(\mu a)^{1-d/2}K_{(d/2)-1}(\mu a)\ .
\end{equation}
It is due solely to the interaction between the two parts of the vector source $J^{\mu}$ and does not involve the tensor source $J^{\mu\nu}$.

	Substituting the sources (\ref{correntetotal}) in the contributions $E_{2}$ and $E_{3}$ exposed in Eq. (\ref{E=E1+E2+E3+E4}), proceeding as in the previous section and using the result (\ref{integral}), we can write
\begin{eqnarray}
\label{asd9}
E_{2}=E_{3}=L^{D}\frac{1}{(2\pi)^{d/2}}\frac{\mu}{2}(w_{\lambda}V^{\lambda}+v_{\lambda}W^{\lambda})\mu^{d-2}(\mu a)^{1-d/2}K_{(d/2)-1}(\mu a)\ .
\end{eqnarray}
The contributions $E_{2}$ and $E_{3}$ come from the interaction between the tensor source $J^{\mu\nu}$ and the vector source $J^{\mu}$.

	With the aid of (\ref{asd7}), (\ref{asd8}) and (\ref{asd9}), using the fact that
\begin{eqnarray}
\frac{\partial}{\partial x}\Bigl(x^{1-(d/2)}K_{(d/2)-1}(x)\Bigr)&=&-x^{1-(d/2)}K_{d/2}(x)\cr\cr
xK_{(d/2)+1}(x)-dK_{d/2}(x)&=&xK_{(d/2)-1}(x)\ ,
\end{eqnarray}
and performing some simple manipulations, we have the total interaction energy density
\begin{eqnarray}
\label{energiaCSKRfinal}
{\cal E}=\frac{E}{L^{D}}&=&\frac{1}{L^{D}}(E_{1}+E_{2}+E_{3}+E_{4})\cr\cr
&=&\frac{1}{(2\pi)^{d/2}}\Bigl[{v_{\lambda}w^{\lambda}}+
\mu(w_{\lambda}V^{\lambda}+v_{\lambda}W^{\lambda})+\mu^{2}V_{\lambda}W^{\lambda}\Bigr]\mu^{d-2}(\mu a)^{1-d/2}K_{(d/2)-1}(\mu a)\cr\cr
&-&\frac{1}{(2\pi)^{d/2}}\Bigl[({\bf V}_{\perp}\cdot{\bf W}_{\perp})K_{d/2}(\mu a)-({\bf V}\cdot{\hat a})({\bf W}\cdot{\hat a})\mu aK_{(d/2)+1}(\mu a)\Biggr]\mu^{d}(\mu a)^{-d/2}\ .
\end{eqnarray}


Notice that, on the contrary to the Kalb-Ramond case, the energy (\ref{energiaCSKRfinal})	exhibits dependence on the perpendicular as well as on the parallel sectors of the vectors $V^{\mu}$ and $W^{\mu}$.

In the most interesting situation we have point-like branes, what corresponds to take $D=0$ and $d=3$, and we must have $v^{\mu}=v\eta^{\mu 0}$ and $w^{\mu}=w\eta^{\mu 0}$. So the energy (\ref{energiaCSKRfinal}) reads
\begin{eqnarray}
\label{energiaCSKRfinald=3}
E&=&\frac{vw}{4\pi}\frac{\exp(-\mu a)}{a}+\frac{\mu}{4\pi}(wV^{0}+vW^{0})\frac{\exp(-\mu a)}{a}
+\frac{\mu^{2}V^{\lambda}W_{\lambda}}{4\pi}\frac{\exp(-\mu a)}{a}
\cr\cr
&-&\frac{\exp(-\mu a)}{4\pi a^{3}}\Bigl[\Bigl(1+\mu a\Bigr)({\bf V}\cdot{\bf W})
-\Bigl(3+3\mu a+\mu^{2}a^{2}\Bigr)({\bf V}\cdot{\hat a})({\bf W}\cdot{\hat a})\Bigr]\ .
\end{eqnarray}

	From the result (\ref{energiaCSKRfinald=3}) we can see that the $\mu$-parameter accomplishes two roles: it acts like a mass for the field ${\cal X}^{\alpha\beta,\gamma}$, producing an exponential decay for the interaction energy between the external sources and, also, produces the interaction between the tensor and vector sectors of the field ${\cal X}^{\alpha\beta,\gamma}$. When $\mu=0$, the above energy reduces to a coulombian interaction between the electric charges $-v$ and $-w$, which comes strictly from the electromagnetic sector, and the interaction between the skew-symmetric rank-2 sources, which comes strictly from the Kalb-Ramond field and exhibits a dipole-like behavior.
		
	If we have only electric charges involved, that is, whenever $V^{\lambda}=W^{\lambda}=0$, the energy (\ref{energiaCSKRfinald=3}) reduces to the Yukawa interaction between two electric charges, $-v$ and $-w$, intermediated by the massive vector field (Proca field).

	A more interesting situation happens if we have only the sources for the Kalb-Ramond field ($v=w=0$). In this case the energy (\ref{energiaCSKRfinald=3}) is the same as the one obtained for the interaction between two point-like dipoles intermediated by a bosonic massive field \cite{BaroneHidalgo,BaroneHidalgo2} plus a Yukawa interaction. So, for point-like sources, in addition to describe a kind of point-like dipole for the skew-symmetric field, the source (\ref{correnteKRgeral}) describes a kind of point-charge for this field as well. This fact becomes clearer if we choose $V^{0}$ and $W^{0}$ in such a way that $V^{\lambda}W_{\lambda}=0$, with $v=w=0$, what brings Eq. (\ref{energiaCSKRfinald=3}) to the form of a pure interaction energy between point-like dipoles intermediated by a massive bosonic field.
	
	Taking ${\bf V}={\bf W}=0$ and $v=w=0$ in Eq. (\ref{energiaCSKRfinald=3}) we have
\begin{equation}
\label{tgb1}
E=\frac{(\mu V^{0})(\mu W^{0})}{4\pi}\frac{\exp(-\mu a)}{a}\ ,
\end{equation}
which is a Yukawa potential between the Kalb-Ramond charges $\mu V^{0}$ and $\mu W^{0}$. If we take, in Eq. (\ref{tgb1}), the limit $\mu\to0$ in such a way that $(\mu V^{0})\to\mbox{finite}$ and $(\mu W^{0})\to\mbox{finite}$, we have the Coulomb potential.

	The same analysis could be done for the energy density (\ref{energiaCSKRfinal}), which is valid for $d=1,2,3$. The second line of (\ref{energiaCSKRfinal}) has the same $a$-dependence as the one exhibited by the interaction energy density between two uniform charge distributions along parallel branes \cite{BaroneHidalgo}. The third line of (\ref{energiaCSKRfinal}) has the same $a$-dependence as the one exhibited by the interaction energy density between two dipole distributions along parallel branes intermediated by a bosonic field \cite{BaroneHidalgo}.
	
	Result (\ref{energiaCSKRfinal}) corroborates the fact that the model (\ref{LCSKR}) is equivalent to a massive vector field, as pointed out in references \cite{CSKR,Lahiri}. However, as long as $\mu\not=0$, our situation is actually different from the purely Proca-like field, even though $\mu\not=0$ corresponds to the exchange of a neutral massive spin gauge boson. This example points out that the particular mass generation mechanism has non trivial consequences on the interaction energy between the skew-symmetric sources and the model described by (\ref{LCSKR}) with the source (\ref{correnteKRgeral}) is richer than the Proca field.

\section{Conclusions and Final Remarks}

	In this paper, we have carried out a study of the interaction between time-independent external sources intermediated by bosonic fields. Special attention has been paid to skew-symmetric currents coupled to the so-called Kalb-Ramond field. The interaction energy between two distributions of generalized dipoles along parallel branes in $3+1$ dimensions intermediated by their coupling to the Kalb-Ramond field has been calculated. As a special result, we have considered the interaction energy between two point-like generalized dipoles.
	
	It has been shown that it is not possible to exist a true point-like static charge or true point-like dipole for the Kalb-Ramond field and the calculation of the interaction energy between two charges distributions along two parallel lines intermediated by this field in a space-time with arbitrary space dimensions has been performed. 
	
	The interaction energies between skew-symmetric charges (or dipole) distributions intermediated by the Kalb-Ramond field has been worked out, and it turns out to exhibit an overall minus sign in comparing with the electromagnetic case. Thus we can say that, for the Kalb-Ramond field, charges distributions with the same sign attract one another as expected for a rank-2 tensor field. 

	It has also been calculated the interaction energy between two external time independent sources distributed along two parallel distinct branes in the Cremmer-Scherk-Kalb-Ramond model, which consists in the interaction between the Kalb-Ramond field and the electromagnetic one and can be seen as an extension to $3+1$ dimensions of the Chern-Simons model. We have considered an external time-independent source composed by two charges distributions for the Maxwell field and two sources for the Kalb-Ramond field placed at two distinct branes. We have shown that, in the case where the branes are point-like, in spite of the considered model to be gauge invariant, when we have only electromagnetic charges we have the Yukawa interaction between them. When we have only skew-symmetric sources, the interaction between them is the one intermediated by a massive bosonic field and exhibits the same behavior of the interaction between two dipoles plus a Yukawa potential.
	
	So, from the charge-dipole interaction, we may conclude that the coupled system Maxwell-Kalb-Ramond field is selected as a way to endow a spin-1 gauge boson with a non-zero mass.
	
	The case of moving sources for the Kalb-Ramond field, which is not considered in this paper, is a relevant and vast theme that deserves to be investigated, also in the context of the Cremmer-Scherk-Kalb-Ramond model. For instance, the case of point-like time dependent sources can be studied with the aid of Eq. (\ref{correntemov}).

\ 

{\bf Acknowledgments}

	The authors would like to thank FAPEMIG, CNPq and CAPES for invaluable financial support.



\end{document}